\documentclass[12pt]{article}

\usepackage{scicite}
\usepackage{color,soul}

\usepackage{times}

\usepackage{epsfig}
\usepackage{graphicx}
\usepackage{amsmath,amssymb}

\newcommand{\beq}[1]{
\begin{equation}
\label{e#1} }

\newcommand{\eeq}{
\end{equation}
}

\newenvironment{sciabstract}{%
\begin{quote} \bf}
{\end{quote}}

\newcounter{lastnote}

\title{Electrical switching of an antiferromagnet}

\author
{Peter Wadley,$^{1,\#,\ast}$ Bryn Howells,$^{1,\ast}$ Jakub \v{Z}elezn\'y,$^{2,3}$ Carl Andrews,$^{1}$ Victoria Hills,$^{1}$ \\ Richard P. Campion,$^{1}$ V\'{\i}t Nov\'ak,$^{2}$ Frank Freimuth,$^{4}$ Yuriy Mokrousov,$^{4}$ \\ Andrew W. Rushforth,$^{1}$ Kevin W. Edmonds,$^{1}$  Bryan L. Gallagher,$^{1}$ Tom\'a\v{s}~Jungwirth$^{2,1}$\\
\\
\normalsize{$^{1}$School of Physics and
Astronomy, University of Nottingham,}\\
\normalsize{Nottingham NG7 2RD, United Kingdom}\\
\normalsize{$^{2}$Institute of Physics ASCR, v.v.i., Cukrovarnick\'a 10, 162 53
Praha 6, Czech Republic}\\
\normalsize{$^{3}$Faculty of Mathematics and Physics, Charles University, Ke Karlovu 3, 121 16 Prague 2, Czech Republic}\\
\normalsize{$^{4}$Peter Gr\"{u}nberg Institut and Institute for Advanced Simulation,}\\
\normalsize{Forschungszentrum J\"{u}lich and JARA, 52425 J\"{u}lich, Germany}\\  
\\
\normalsize{$^\ast$ Authors contributed equally to this work.}
\\
\normalsize{$^\#$To whom correspondence should be addressed; E-mail:  Peter.Wadley@nottingham.ac.uk.}
}

\date{}
\begin{document} 


\baselineskip24pt


\maketitle 

\begin{sciabstract}
 
Louis N\'eel pointed out in his Nobel lecture that while abundant and interesting from a theoretical viewpoint, antiferromagnets did not seem to have any applications. Indeed, the alternating directions of magnetic moments on individual atoms and the resulting zero net magnetization make antiferromagnets hard to control by tools common in ferromagnets. Remarkably, N\'eel in his lecture provides the key which, as we show here, allows us to control antiferromagnets by electrical means analogous to those which paved the way to the development of ferromagnetic spintronics applications.  The key noted by N\'eel is the equivalence of antiferromagnets and ferromagnets for effects that are an even function of the magnetic moment. Based on even-in-moment  relativistic transport phenomena, we demonstrate room-temperature electrical switching between two stable configurations  combined with electrical read-out in antiferromagnetic CuMnAs thin film devices. Our magnetic memory is insensitive to and produces no magnetic field perturbations which illustrates the unique merits of antiferromagnets for spintronics. 
\end{sciabstract}

Information technology devices are either charge-based or spin-based. Among the weaknesses of charge-based devices is that perturbations such as ionizing radiation can lead to uncontrolled charge redistributions and, as a consequence, to data loss. All commercial spin-based devices rely on one principle in which the opposite magnetic moment orientations in a ferromagnet (FM) represent the zeros and ones \cite{Chappert2007}. This technology is behind memory applications ranging from kilobyte magnetic stripe cards to megabyte magnetoresistive random access memories (MRAMs) and terabyte computer hard disks. Since based on spin, the devices are robust against charge perturbations. Moreover, they are non-volatile and, compared to the charge flash memory, MRAM offers short read/write times suitable for the main random-access computer memories. However, the FM moments can be unintentionally reoriented and the data erased by disturbing magnetic fields generated externally or internally within the memory circuitry. The vision behind our work involves information technologies based on antiferromagnets (AFMs) that are robust against charge and magnetic field perturbations, and  which utilize other advantages of AFMs compared to FMs including the ultrafast magnetic dynamics  and the broad range of metal, semiconductor, or insulator materials with room-temperature AFM order \cite{Kimel2004,Fiebig2008,Yamaoka1974,Zhang2014e,Jungwirth2011,Maca2012}. 

Several roots from micro-device, relativistic quantum physics, and materials research had to meet to provide us with the means to experimentally demonstrate the concept of a room-temperature AFM memory with electrical writing and reading. In FMs the bistability, i.e., the energy barrier separating two stable directions of ordered spins is due to the magnetic anisotropy energy. It is an even function of the magnetic moment which, following N\'eel's general principle noted in his Nobel lecture \cite{Neel1970}, implies that magnetic anisotropy is readily present also in AFMs~\cite{Umetsu2006}. Bistability can be realized in classes of AFMs which possess biaxial magnetic anisotropy. 

The magneto-transport counterpart of magnetic anisotropy is anisotropic magnetoresistance (AMR). In the early 1990's, the first generation of FM MRAM micro-devices used AMR for the electrical read-out of the memory state \cite{Daughton1992}. AMR is an even function of the magnetic moment which, following again N\'eel's principle, implies its  presence  in AFMs \cite{Shick2010}. While  AMR in AFMs  was experimentally confirmed in several recent studies \cite{Park2011b,Marti2014,Fina2014}, efficient means for manipulating the AFM moments have remained elusive. External magnetic field, whose coupling is linear in the magnetic moment,  favors parallel alignment of the moments, i.e., acts against the staggered exchange field in the AFM. 

On the other hand, current-induced even magnetic torques of the form $d{\bf M}/dt\sim {\bf M}\times({\bf M}\times{\bf p})$,  which are used for electrical writing in the most advanced FM spin-transfer-torque MRAMs \cite{Chappert2007}, have been proposed to allow for a large angle reorientation of the AFM moments \cite{Gomonay2010}. Here ${\bf M}$ is the magnetic moment vector and ${\bf p}$ is the electrically injected carrier spin-polarization.  Translated to AFMs, the effective field proportional to $({\bf M}_{A,B}\times{\bf p})$ which drives the torque $d{\bf M}_{A,B}/dt\sim {\bf M}_{A,B}\times({\bf M}_{A,B}\times{\bf p})$ on individual spin sublattices A and B is staggered, i.e., alternates in sign between the opposite spin sublattices. The staggered property of the field is the key for its strong coupling to the N\'eel order. 

In FM spin-transfer-torque MRAMs, spin polarized carriers are injected into the free FM layer from a fixed FM polarizer by an out-of-plane electrical current driven through the FM-FM stack. Reversible 180$^\circ$ switching is achieved by reversing the polarity of the out-of-plane current. In analogy with FM spin-transfer-torque MRAMs, the above theory proposed for AFMs \cite{Gomonay2010} assumes injection of the spin polarized carriers into the AFM from a fixed FM polarizer by out-of-plane electrical current driven in a FM-AFM stack. Unlike the FM-FM stack, the effect of the spin current injected from a FM to an AFM is independent of the current polarity. The AFM spin-axis can in principle be switched from the parallel to the perpendicular direction with respect to the magnetization of the FM polarizer but cannot be switched back electrically \cite{Gomonay2010}. Another limitation of the stack geometry with out-of-plane spin injection is that the absorption of carrier spins is limited to magnetic film thicknesses of the order of the spin diffusion length which in AFMs is typically on the nanometer scale \cite{Acharyya2011}. 

Several of us have predicted \cite{Zelezny2014} that relativistic quantum mechanics may offer staggered current induced fields which do not require external polarizers and act in bare AFM films.  Instead of changing the current polarity, a reversible 90$^\circ$ switching is facilitated by applying in-plane electrical currents along two orthogonal directions. The effect occurs in AFMs with specific crystal and magnetic structures for which the spin sublattices form space-inversion partners. Among these materials is a high N\'eel temperature AFM, tetragonal-phase CuMnAs, which we have recently synthesized in the form of single-crystal epilayers structurally compatible with common semiconductors, such as Si or GaP \cite{Wadley2013}. Relativistic current-induced fields strongly coupled to the N\'eel order, combined with AFM CuMnAs epilayers and micro-device AMR, are the key elements that led to our demonstration of the electrical switching in an AFM memory resistor which we now describe in more detail.

We first outline the basic theoretical considerations behind our experiments. Relativistic current-induced fields observed previously in broken inversion-symmetry FM crystals \cite{Bernevig2005c,Chernyshov2009,Endo2010,Fang2011,Manchon2008,Miron2010,Pi2010,Suzuki2011,Miron2011b} can originate from the inverse spin galvanic effect \cite{Silov2004,Kato2004b,Ganichev2004b,Wunderlich2004,Wunderlich2005} which is illustrated in Fig.1~A. For illustrative purposes we consider here that the spin-orbit term in the Hamiltonian of the system, which originates from broken inversion symmetry in the crystal, has a two-dimensional Rashba form with carrier spins aligned in the direction perpendicular to the momentum. The applied electrical current induces an asymmetric non-equilibrium  distribution function of carrier states and, as a result, a non-zero carrier spin-polarization ${\bf p}$ aligns perpendicular to the applied current. In FMs, the non-equilibrium carrier spin density acts on  magnetic moments as an effective magnetic field when carrier spins are exchange-coupled to the magnetic moments. The resulting field-like torque has a form  $d{\bf M}/dt\sim {\bf M}\times{\bf p}$. When the sense of the inversion symmetry of the lattice reverses, the sign of the carrier spin polarization and of the corresponding effective field also reverses for the same orientation of the applied current, as illustrated in Fig.~1B. 

The full lattice of our CuMnAs crystal, shown in Fig.~1C,  has an inversion symmetry; the center of inversion of the lattice is at an interstitial position, highlighted by the green ball in the figure.  This implies that  the mechanism described in Figs.~1A,B will not generate a net current-induced spin-density when integrated over the entire crystal. However, Mn atoms  form  two sublattices depicted in Fig.~1C in red and purple  whose local environment has broken inversion symmetry and the two Mn sublattices form inversion partners. The inverse spin galvanic mechanisms of Figs.~1A,B will generate locally non-equilibrium spin polarizations  of opposite signs on the inversion-partner Mn sublattices. For these staggered fields to couple strongly to the N\'eel order it is essential that the inversion-partner Mn sublattices coincide with the two spin-sublattices A and B of the AFM ground-state  \cite{Zelezny2014}. The resulting spin-sublattice torques have the form  $d{\bf M}_{A,B}/dt\sim {\bf M}_{A,B}\times{\bf p_{A,B}}$ where, in analogy with the even torques, the effective field proportional to ${\bf p_{A}}=-{\bf p_{B}}$ acting on the spin-sublattice magnetizations alternates in sign between the two sublattices. This again implies the strong coupling to the N\'eel magnetic order. As shown in Fig.~1C, CuMnAs crystal and magnetic structures fulfil these symmetry requirements \cite{Wadley2013}. 

To quantitatively estimate the strength of the staggered current induced field we performed microscopic calculations based on the Kubo linear response formalism. The electronic structure of CuMnAs was obtained by the J\"{u}lich density functional theory 
code \texttt{FLEUR} which is an implementation of the full-potential linearized augmented plane-wave method \cite{Freimuth2014a}. In Fig.~1D we plot the resulting  components of the in-plane current induced field transverse to the magnetic moments at spin-sublattices A and B as a function of the in-plane magnetic moment angle $\varphi$ measured from the x-axis ([100] crystal direction). The electrical current of 10$^7$~Acm$^{-2}$ is applied along the x-axis  or the y-axis. 

The calculations confirm the desired opposite sign of the current induced field on the two spin-sublattices. They also highlight the expected dependence on the magnetic moment angle which implies  that the AFM moments will tend to align perpendicular to the applied current. For reversible electrical switching between two stable states, the setting current pulses can therefore be applied along  two orthogonal in-plane cubic axes of CuMnAs. We also emphasize that the amplitude of the effect seen in Fig.~1D is comparable to typical current induced fields applied in FMs, suggesting that CuMnAs is a favorable material for observing the effect in an AFM.

For our experiments we prepared an epitaxial $46$-nm thick film of CuMnAs which is a member of a broad family of high-temperature I-Mn-V AFM compounds \cite{Jungwirth2011,Maca2012,Wadley2013}. Our tetragonal CuMnAs epilayer, whose transmission electron microscopy image is shown in Fig.~2A,  is grown on a lattice-matched GaP(001) substrate and buffer layers and exhibits excellent crystal quality, chemical order and compatibility with semiconductor growth and microfabrication technologies \cite{Wadley2013}. The room-temperature electrical resistivity is around 160 $\mu\Omega$ cm. Neutron diffraction confirmed collinear AFM order with a N\'eel temperature $T_N=480$~K \cite{Wadley2013,Hills2015}. 

Measurements were performed at room temperature on a set of devices of the geometry shown in Fig.~2B. Current pulses for electrical switching of the AFM states were applied along the principal arms 1-3-5-7, which are of width either 8$\mu$m or 28$\mu$m in the two types of devices studied. The arms along the diagonal axes 2-4-6-8 are used for electrical monitoring of the AFM states, and are of width 8$\mu$m in both types of device. Due to current spreading, the current density in the central region of the devices is smaller than in the arms. From finite element modelling, the current density at the center is estimated to be 30\% of the current density in the arms for the 8$\mu$m device, and 60\% for the 28$\mu$m device. 

In Figs.~2C,D we present measurements on the 8~$\mu$m device with a 100~$\mu$A probe current and with setting 50~ms current pulses of amplitude 50~mA, corresponding to a current density of around 4$\times 10^6$~Acm$^{-2}$ in the center of the device.
Measurements of the electrical resistivity during and after the current pulse indicate that the device temperature increases by up to 60K due to Joule heating. A delay of 5 minutes between setting and probing the state ensures that the device is at thermal equilibrium during probing. Reproducible switching is observed in the detected transverse resistance signal measured by voltage contacts 2-6 with the probe current along the [110] axis applied between current contacts 4-8.  The setting current pulses are applied in these measurements successively along the [100] axis (current contacts 1-5) and [010] axis (current contacts 3-7). As discussed below, the observed electrical signal shows the expected symmetry of transverse AMR (also known as the planar Hall effect) induced by N\'eel order current-induced torques. The panels 2C and 2D only differ in the polarity of the setting pulses and confirm the expected independence of the current-induced switching on the polarity of the applied current. 

In Fig.~3 we demonstrate that the probe electrical signals are consistent with the AMR symmetries. Side by side we compare the transverse AMR signals, $\Delta R_t/\bar{R}_{sq}$,  and longitudinal AMR signals, $\Delta R_{sq}/\bar{R}_{sq}$. Here $\Delta R_{t(sq)}$ is the difference between the transverse (sheet) resistance after the setting current pulse and average value, and $\bar{R}_{sq}$ is the average sheet resistance. Each row in Fig.~3 corresponds to a different axis along which we apply the probe current. From top to bottom, the probe current is applied along the crystal axis [110] (contacts 4-8), [1$\bar{1}$0] (contacts 2-6), [100] (contacts 1-5), and [010] (contacts 3-7). The setting current pulses are applied successively along the [100] and [010] axes in all measurements in Fig.~3. Consistent with the AMR phenomenology, the transverse AMR signals are detected for the AFM spin-axes angle set towards $\pm 45^\circ$ from the probe current and the transverse AMR flips sign when the probe current is rotated by $90^\circ$. The corresponding longitudinal AMR signals vanish in this geometry. For AFM spin-axes set towards $\pm 90^\circ$ from the probe current the transverse and longitudinal AMR signals switch places compared to the $\pm 45^\circ$ angle between the probe-current and spin-axis. This is again consistent with the AMR symmetries, and the AMR nature of the electrical signals is further confirmed by comparable amplitudes of the transverse and longitudinal signals. All samples studied showed similar switching characteristics.

A more detailed study of the dependence of the switching on the length and magnitude of the electrical excitation is presented in Figs.~4A,B. Measurements in panel 4A are performed on the 8~$\mu$m device with the setting 50~ms 50~mA pulses applied successively in the same direction at five minute intervals. For a fixed pulse direction, the amplitude of the probe signal initially increases and then tends to saturate with increasing pulse number towards the low/high value. We interpret this observation that successively more AFM domains in the sample are oriented by the current pulses in the same direction, determined by the direction of the setting current. The $\sim10~\mu$m width of the central section sets an upper limit on the domain size. When reorienting the setting current by 90$^{\circ}$ the trend immediately reverses, showing again the initial change in the probe signal followed  by the saturation towards the high/low value. The reversed trend is observed only when the direction of the setting current is switched by 90$^{\circ}$.

Fig.~4B shows the dependence of the switching signal on the amplitude of a 50~ms pulse varied between 70 and 120~mA which in the measured 28~$\mu$m device corresponds to $3-6\times 10^6$~Acm$^{-2}$ in center of the device. We observe no effect of the applied setting current below a threshold of $\sim 6\times 10^6$~Acm$^{-2}$, above which the switching signal starts to increase with the current amplitude. 

From the series of electrical measurements discussed above we conclude that the electrical writing and reading of our CuMnAs AFM memory is robust, reproducible, and requires comparable current excitations to FM devices. The switching currents applied are significantly lower than in the first observation of spin-orbit torque switching in ferromagnetic metals, where 100~MA~cm$^{-2}$ pulses were used to reverse magnetization in Pt/Co bilayer \cite{Miron2011b}. In Figs.~4C,D we illustrate the unique characteristics of the AFM memory.  Fig.~4C compares the behavior of the 28~$\mu$m  device at zero external magnetic field and at the largest field of 4.4~kOe available in our measurement system. The delay between the 50~ms 110~mA pulses (current density $8\times 10^6$~Acm$^{-2}$) and the probe in these measurements is 30 minutes and the time gap between cycling the memory device without and with magnetic field is 24 hours. The magnetic field was applied in-plane and during both pulsing and probing the AFM memory. We do not observe any change in the behavior of our magnetic memory without and with the applied magnetic field.  Indeed our CuMnAs is a fully compensated AFM which is insensitive to magnetic field perturbations and generates negligible stray magnetic fields. This is confirmed by the SQUID magnetization measurements in Fig.~4D showing only the diamagnetic background of the sample substrate. These characteristics make the AFM memory fundamentally distinct from FM devices.

To conclude, reorienting AFMs with an efficiency comparable to FMs requires a field compatible with the N\'eel magnetic order. Generating external magnetic fields with a sign alternating on the scale of a lattice constant at which moments alternate in AFMs has been considered unfeasible, hindering the applications of these abundant magnetic materials. We have succeeded  in generating electrically such an alternating field and utilized its strong coupling to the N\'eel order for switching the CuMnAs AFM between two  configurations which are stable at ambient conditions and robust against magnetic field perturbations. The staggered field is generated electrically with comparable efficiency and by relativistic physics analogous to that which has driven recent major advancements in the FM spintronics research. 

The staggered current induced fields we observe are not unique to CuMnAs. A high N\'eel temperature AFM Mn$_2$Au \cite{Barthem2013} is another example in which the spin-sublattices form inversion partners and where theory predicts large torques of the form $d{\bf M}_{A,B}/dt\sim {\bf M}_{A,B}\times{\bf p_{A,B}}$ with  ${\bf p_{A}}=-{\bf p_{B}}$ \cite{Zelezny2014}. From microscopic density-functional calculations, we obtain a current-induced field of around 20~Oe per 10$^7$ Am$^{-2}$ in Mn$_2$Au, which combined with its higher conductivity may make this a favorable system for observing current-driven AFM switching. AFMs which do not possess these specific symmetries can be alternatively interfaced with spin-orbit coupled non-magnetic (NM) layers. A spin current will be injected into the AFM from the NM layer by an applied in-plane electrical current via the spin Hall effect, generating the spin torque $d{\bf M}_{A,B}/dt\sim {\bf M}_{A,B}\times({\bf M}_{A,B}\times{\bf p})$ \cite{Zelezny2014}. The same symmetry torque can be also generated by the spin-orbit Berry-curvature mechanism acting at the inversion-asymmetric AFM/NM interface \cite{Zelezny2014}. The in-plane current geometry allows for applying the setting currents again along different axes to realize reversible electrical switching. Our experiments in CuMnAs combined with the prospect of other realizations of these relativistic non-equilibrium phenomena in AFMs brings us to the conclusion that AFMs are now ready to join the rapidly developing fields of basic and applied spintronics and to enrich this area of solid state physics and microelectronics by the range of unique characteristics of AFMs.

\bibliographystyle{Science}

We would like to thank Jasbinder Chauhan for assistance with the microfabrication, and Christopher Nelson for STEM measurements. We acknowledge support from EU ERC Advanced Grant No. 268066, from the Ministry of Education of the Czech Republic Grant No. LM2011026, from the Grant Agency of the Czech Republic Grant no. 14-37427, from the UK EPSRC Grant No. EP/K027808/1, from HGF Programme VH-NG 513 and DFG SPP 1568, and supercomputing resources at J\"ulich Supercomputing
Centre and RWTH Aachen University.

\begin{figure}[h]
\begin{center}
\hspace*{-0cm}\includegraphics[scale=.8]{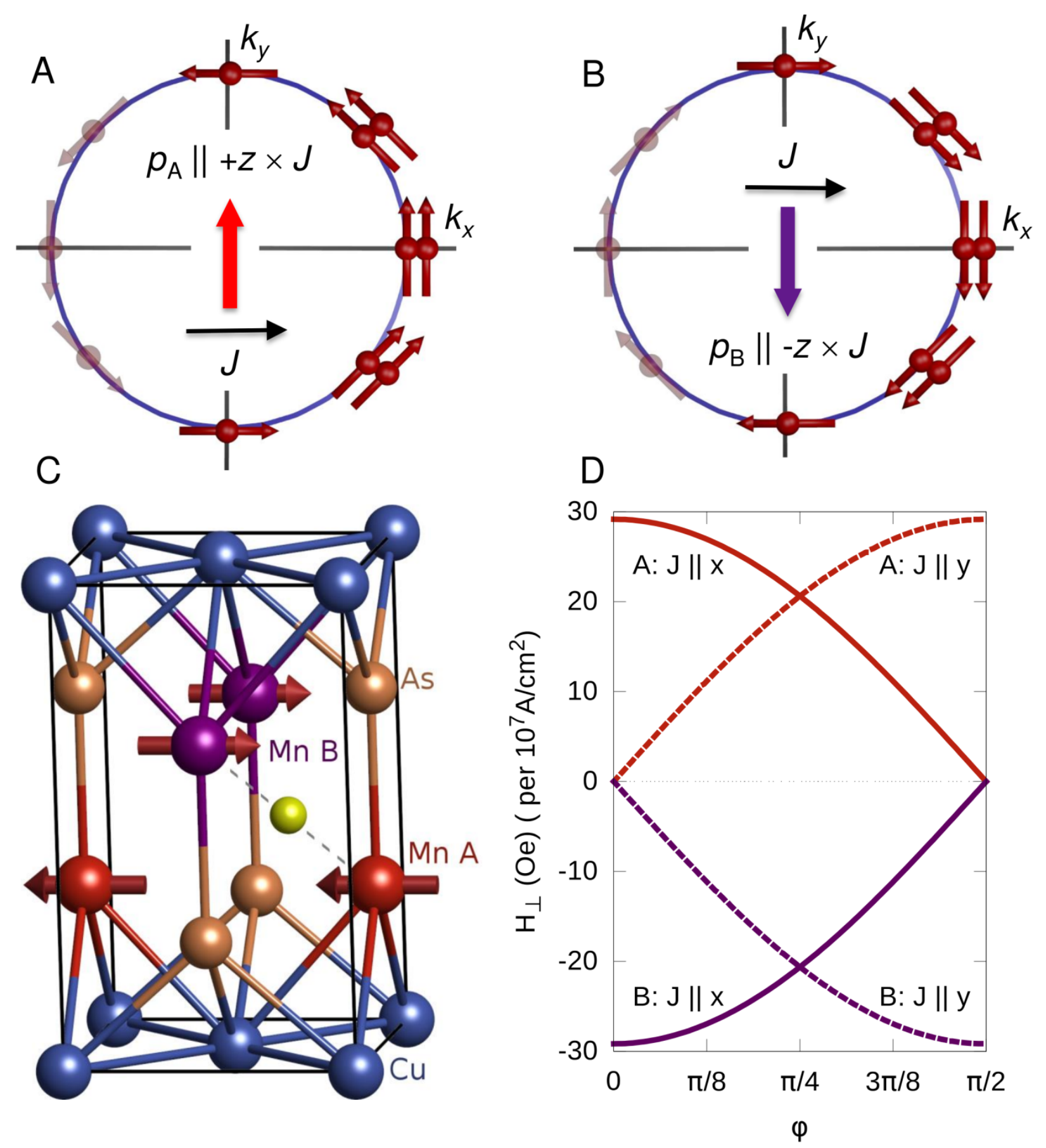}
\end{center}
\caption{(A) Schematic of the inverse spin galvanic effect in a model inversion asymmetric Rashba spin texture (red arrows). $k_{x,y}$ are the in-plane momentum components. The non-equilibrium redistribution of carriers from the left side to the right side of the Fermi surface results in a net in-plane spin polarization (thick red arrow) along $+z\times J$ direction, where $J$ is the applied current (black arrow). (B) Same as (A) for opposite sense of the inversion asymmetry resulting in a net in-plane spin polarization (thick purple arrow) along $-z\times J$ direction. (C) CuMnAs crystal structure and antiferromagnetic ordering. The two Mn spin-sublattices A and B (red and purple) are inversion partners. This and panels (A) and (B)  imply opposite sign of the respective local current induced spin polarizations, $p_A=-p_B$, at spin-sublattices A and B.  The full CuMnAs crystal is centrosymmetric around the interstitial position highlighted by the green ball. (D) Microscopic calculations of the components of the current induced field transverse to the magnetic moments at spin-sublattices A and B as a function of the magnetic moment angle $\varphi$ measured from the x-axis ([100] crystal direction). The electrical current is applied along the x and y-axes.}
\label{f1}
\end{figure}

\begin{figure}[h]
\begin{center}
\hspace*{-0cm}\includegraphics[scale=1]{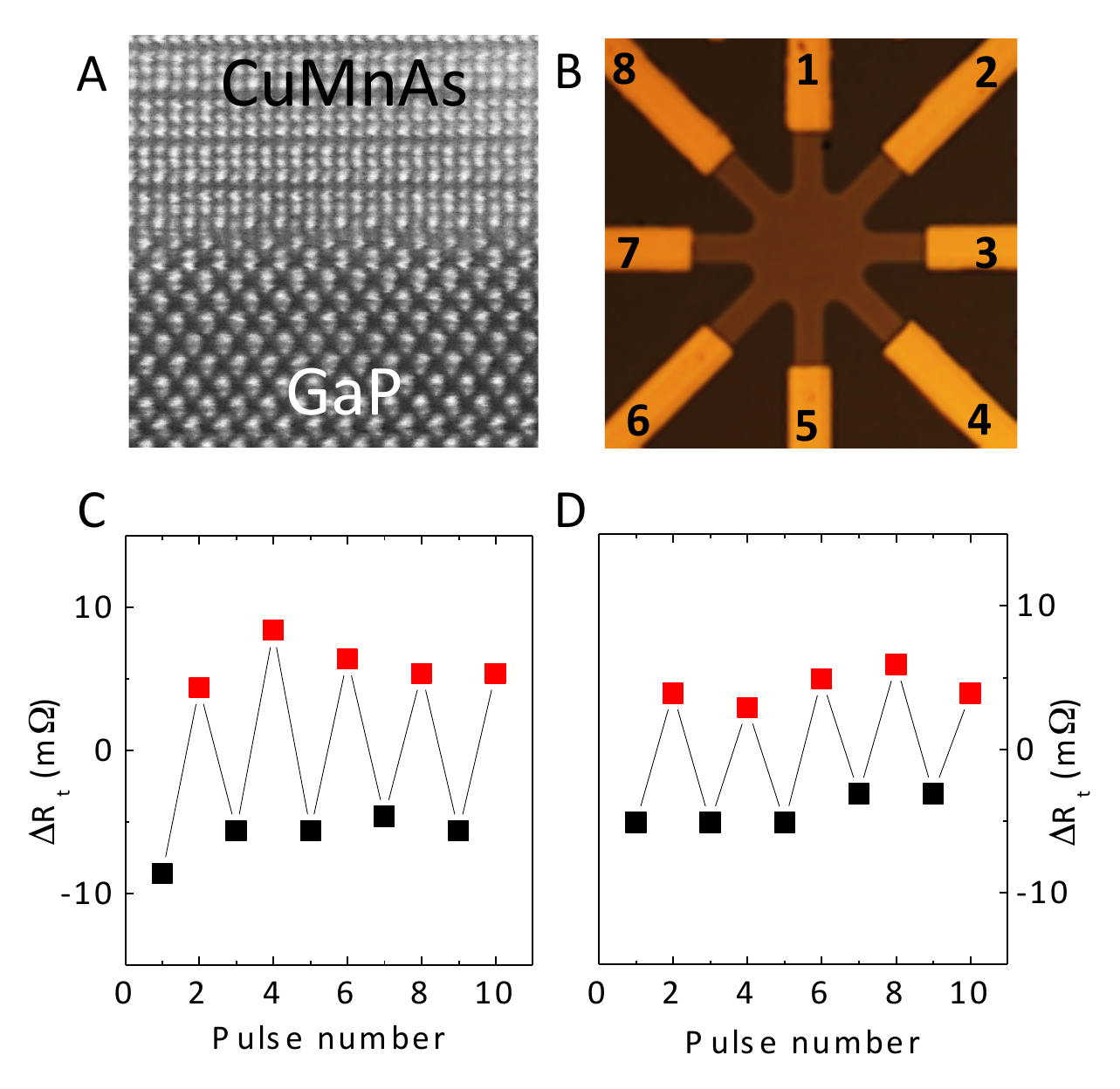}
\end{center}
\caption{(A) Scanning transmission electron microscopy image of CuMnAs/GaP in the [100]--[001] pane. (B) optical microscopy image of device with 8~$\mu$m wide arms. (C) Change in the transverse resistance after applying 50~mA setting current pulses alternately along 1-5 (black) and 3-7 (red). (D) as for (C) but with opposite polarity setting pulse along 5-1 and 7-3.
}
\label{f2}
\end{figure}

\begin{figure}[h]
\begin{center}
\hspace*{0cm}\includegraphics[scale=.7]{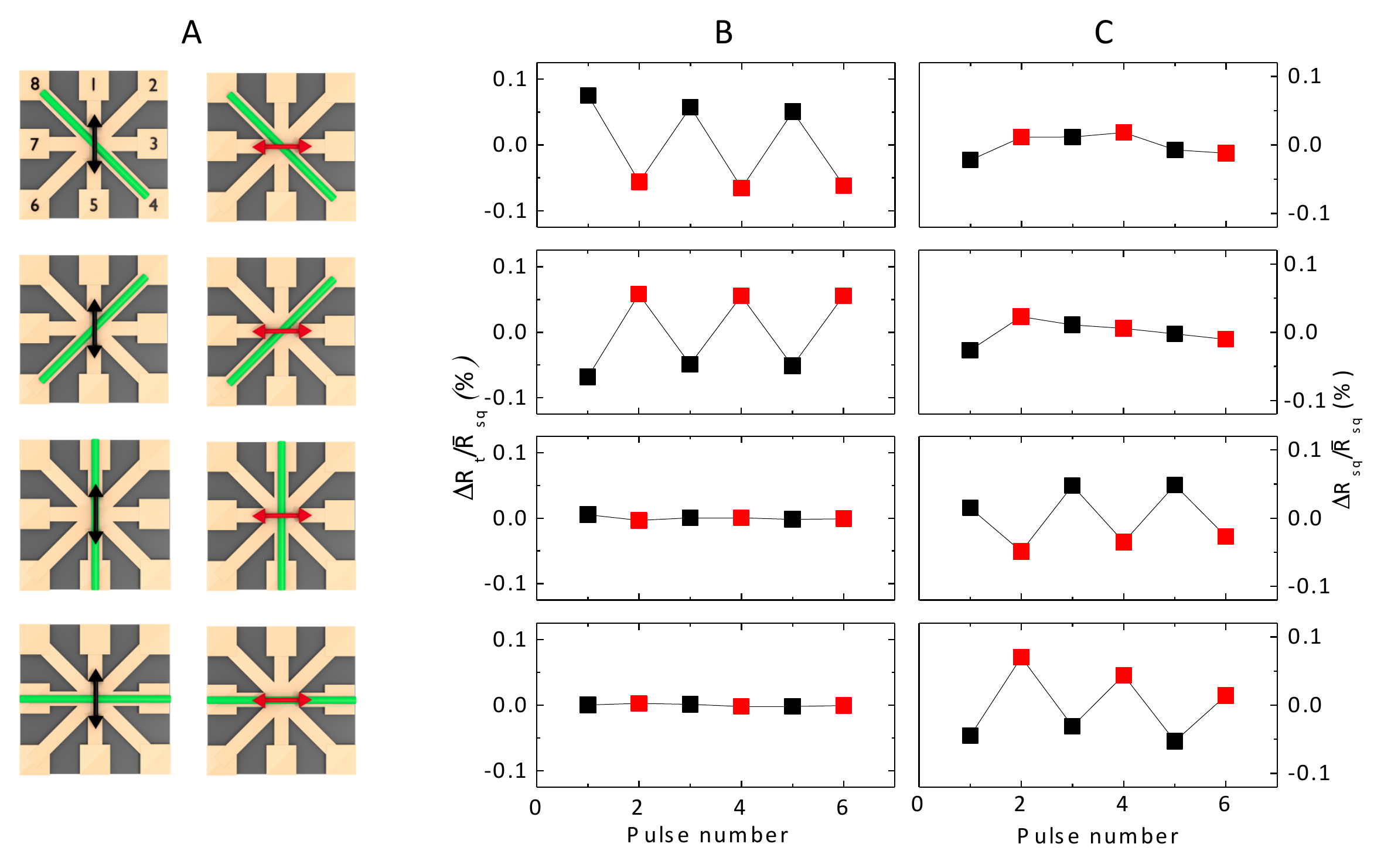}
\end{center}
\caption{(A) Schematic showing the probe current directions (green) and the inferred spin axis (red/black). (B) Normalised transverse resistance change after setting current pulse 1-5 (red) and 3-7 (black) for the probe current directions shown in (A) in a 28~$\mu$m device. (C) As for (B) but for the normalised longitudinal resistance change.
}
\label{f3}
\end{figure}

\begin{figure}[h]
\begin{center}
\hspace*{-4cm}\includegraphics[scale=.8]{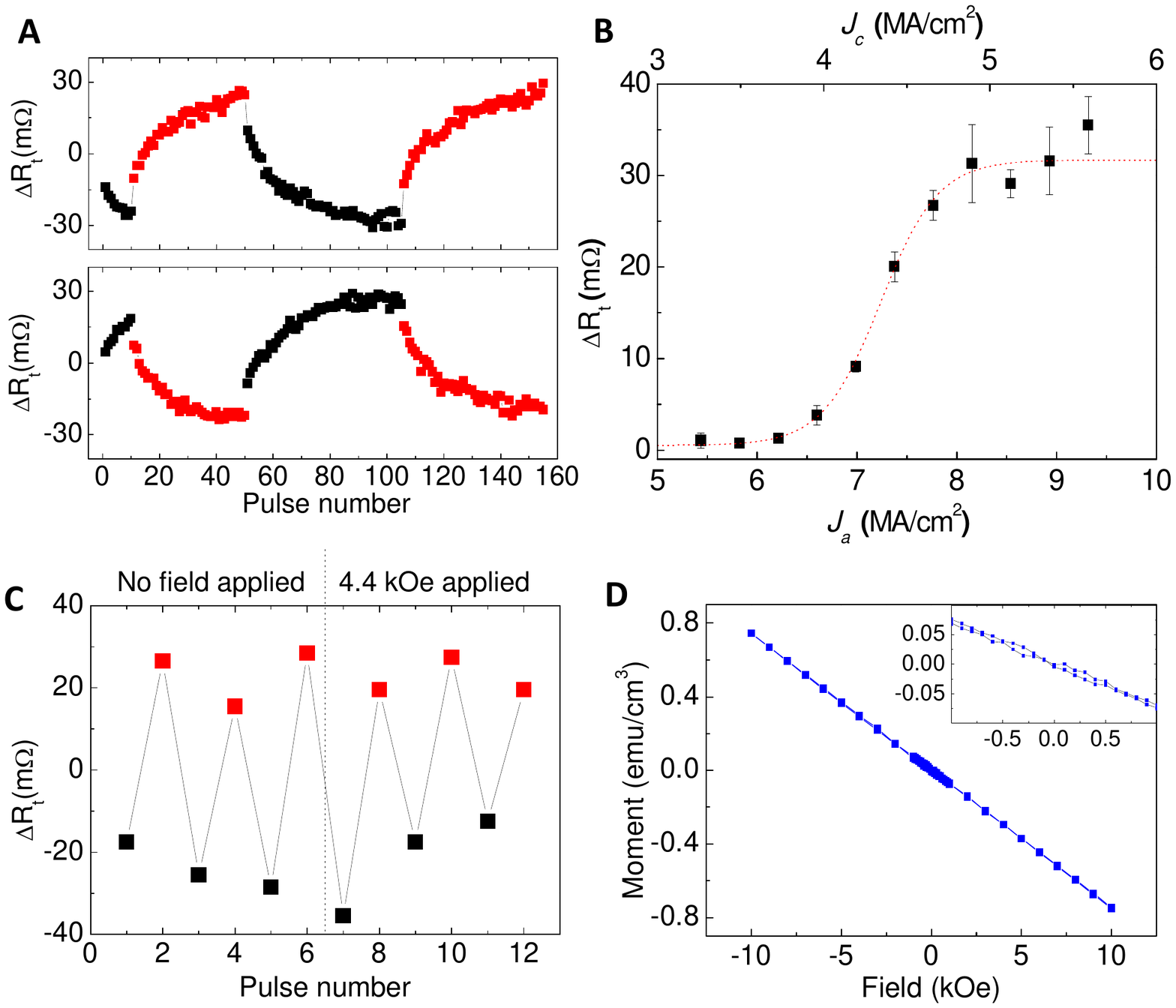}
\end{center}
\caption{(A) Transverse resistance change after successive pulse along 1-5 (black) and 3-7 (red) using probe current along 4-8 (top panel) and 2-6 (bottom panel) in an 8~$\mu$m device. (B) Dependence of the change in transverse resistance on the current density of the setting pulse in the device arms ($J_a$, lower axis) and in the central region ($J_c$, upper axis) in a 28~$\mu$m device. (C) Change in the transverse resistance after current pulses alternating along orthogonal directions without field (pulse number 1--6) and with 4.4~kOe field applied (pulse number 7--12) in the same device as (B). (D) Magnetic moment versus applied field of an unpatterned piece of the CuMnAs/GaP wafer measured by SQUID magnetometer.
}
\label{f4}
\end{figure}
\end{document}